# The Cost Function of a Two-Level Inventory System with identical retailers benefitting from Information Sharing

Amir Hosein Afshar Sedigh[1], Rasoul Haji[2], Seyed Mehdi Sajadifar[2]

[1]Research and Development, Behpardaz Hamrah Samaneh Aval, Tehran, Iran

Email: aafsharsedigh@gmail.com

[2]Industrial Engineering Department, University of Science and Culture, Tehran, Iran

**Abstract**

This paper investigates a two-echelon inventory system with a central warehouse and $N$ ($N >$ 2) retailers managed by a centralized information sharing mechanism. In particular, the paper mathematically models an easy to implement inventory control system that facilitates making use of information. Some assumptions of the paper include: a) constant delivery time for retailers and the central warehouse and b) Poisson demand with identical rates for retailers. The inventory policy comprises of continuous review $(R, Q)$-policy on part of retailers and triggering the system with $m$ batches (of a given size $Q$) at the central warehouse. Besides, the central warehouse monitors retailers' inventory and may order batches sooner than retailers' reorder point, when their inventory position reaches $R + s$. An earlier study proposed the policy and its cost function approximation. This paper derives an exact mathematical model of the cost function for the aforementioned problem.

## 1. Introduction

Improving flexibility and approachability is essential in uncertain and competitive markets. As such, supply chains (SCs) have put more concentration on information technology to distribute benefits among the members (Zhou & Benton, 2007). No information sharing incentivizes individual members to maximize their own benefit. Therefore, customer demand variability is strengthened in upstream levels (e.g. manufacturers and vendors). This leads to a phenomenon called the *bullwhip effect*, originally reported by Forrester (1958) as "demand amplification". The *bullwhip effect* was studied by many researchers, such as Simchi-Levi et al. (2008) and Li & Simchi-Levi (2020). Not only information sharing in a centralized decision-making system weakens demand variability in upstream levels, but it also has benefits such as a fair distribution of profits among all members (Li & Wang, 2007).

Several researchers experimentally and theoretically studied benefits of information sharing in SCs. For instance, Cho & Lee (2013) studied a seasonal SC including one supplier and one retailer. The study indicated that information sharing is beneficial if the lead time is shorter than the seasonal period. Liu et al. (2015) examined the impact of information sharing and process coordination on logistic outsourcing in China. This study showed that these integration mechanisms, especially information sharing, are beneficial. Chen et al. (2007) found full information sharing scenario as the best seceanrio by investigating usage of eight scenarios regarding sharing different types of information. Janaki et al, (2018) indicated the value of information sharing for improving SC performance under uncertainty. However, Ojha et al (2019) indicated the importance of careful choice of type of shared information, when some types of information are not shared for managerial concerns. Liu et al. (2020) indicated that sharing basic inventory information, disregarding production capacity and resource constraints, has the highest impact on the coordination in a decentralize supply chain.

Li & Wang (2007) investigated such problems under three policies, namely installation stock, echelon stock, and information sharing. The study indicated that installation stock policies are easy to operate but neglect performance optimization for not utilizing information about the customers – see Axsäter (1990, 1993, 2000), Forsberg (1995, 1997), Simchi-Levi & Zhao (2007, 2012) for various approaches of using installation stock policies. In echelon stock policies, information on the cumulative inventory positions of all downstream installation is available. Chen & Zheng (1994, 1998) evaluated $(R, nQ)$ echelon stock policies in serial inventory systems and its extension for a multistage inventory system with compound Poisson demand. For the latter they evaluated optimum boundaries and proposed an algorithm for near optimal solution that can also be used for exact solution. It is worth noting that Axsäter & Rosling (1993) indicated that echelon stock policy may outperform the installation stock policy.

In information sharing policy, all or part of information about the inventory position, bill of materials (BOM), etc. is shared among members. This can enhance SC performance due to the reduction of the forecast errors. The information sharing not only helps supply managers to secure transparency and accessibility of information, but it also facilitates decision-making by providing more accurate information about the chain (Simchi-Levi et al. 2008).

In the light of these studies, some studies proposed policies that benefited from shared information. Here we state policies for two-level SC with stochastic demands. For instance, Moinzadeh's (2002) policy is an easy to implement information sharing-based approach for a two-echelon SC with a product, a central warehouse, and some identical retailers under a stationary random demand. He derived an approximate cost function and proposed a heuristic optimization procedure to obtain a near-optimal policy. However, the optimization procedure could not identify the optimal boundaries. Sajadifar & Haji (2007) derived an exact cost function for Moinzadeh's (2002) policy, considering one retailer. They used Axsäter's (1990) installation stock model for one-for-one policy, to evaluate the cost function. Later, Haji & Sajadifar (2008) derived the optimal boundaries for their proposed model in order to have less computational efforts in obtaining the optimal solution. A further development is due to Axsäter & Marklund (2008) who proposed a dynamic information sharing policy for non-identical retailers, that is hard to implement because of its non-static nature. More recently,

Afshar Sedigh et al. (2019) derived the exact cost function and optimal boundaries of Moinzadeh's (2002) policy, for two retailers.

This paper reinvestigates Moinzadeh's (2002) policy by obtaining the exact cost function and optimal boundaries for more than two retailers. This study similar to Moinzadeh (2002) considers some retailers and a central warehouse who have constant delivery times. Retailers face independent Poisson demand with identical rates. The continuous review $(R, Q)$-policy is applied by the retailers. The system triggers with $m$ batches of a given size $Q$ at the central warehouse. When a retailer's inventory position reaches $R + s$, the central warehouse places an order to an outside supplier. This paper derives the exact cost function for different retailers, based on which the optimal policy can be determine. The system is split into some simpler states by conditioning. Then, the exact cost function is obtained for each state. A brief schema of the presented problem is illustrated in Figure 1, where the notations immediately follow in Section 2.

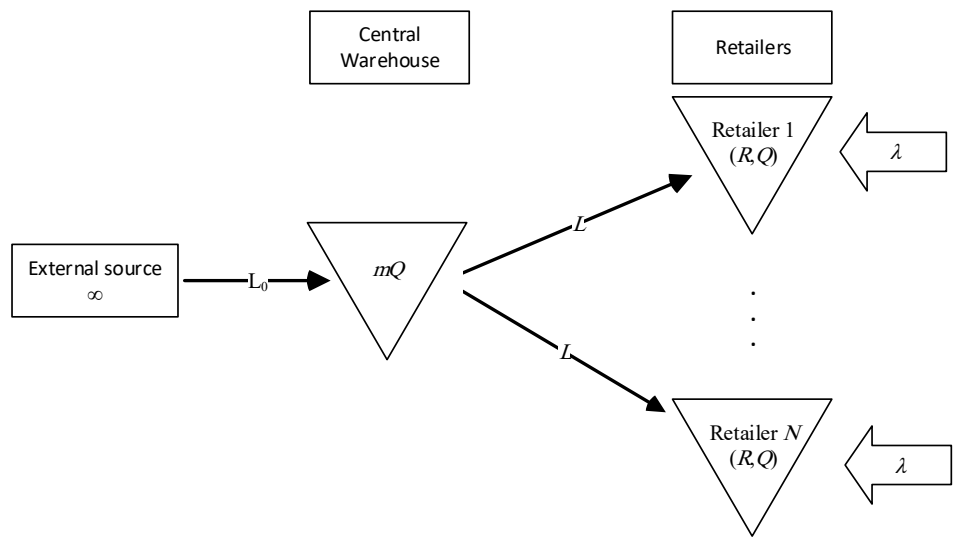

*Figure 1. System schema*

## 2. Problem definition and notations

This section presents employed notations and assumptions. Note that we utilize similar notation to earlier studies for ease of researchers familiar with the field.

### 2.1. Notations

The notations used in this paper are as follows:

$L$: Delivery time from the central warehouse to each retailer

$L_0$: Delivery time from the external supplier to the central warehouse

$\lambda$: Demand intensity at each retailer

$h$: Holding cost per unit per unit time for each retailer

$h_0$: Holding cost per unit per unit time at the central warehouse

$\beta$: Shortage cost per unit per unit time for each retailer

$Q$: Given system's order quantity

$R$: Reorder point for each retailer

$m$: Number of initial batches allocated to the central warehouse

$A_0(t)$: Set of retailers whose inventory positions are less than or equal to $R + s$ at $t$

$A_1(t)$: Set of retailers whose inventory positions are more than $R + s$ at $t$

Moreover, similar to Axsäter (1990), the following notations are used:

$\gamma(S_0)$: Average holding cost in the central warehouse per unit when the inventory position at warehouse is $S_0$

$\Pi^S(S_0)$: Average holding and shortage costs in a retailer when inventory positions of the central warehouse and the retailers are $S_0$ and $S$, respectively[1].

---

[1] Note that, using Lemma 1 and 2 in Forsberg (1995), one can replace $S_0$ and $S$ by $k$ and $R + j$, respectively.

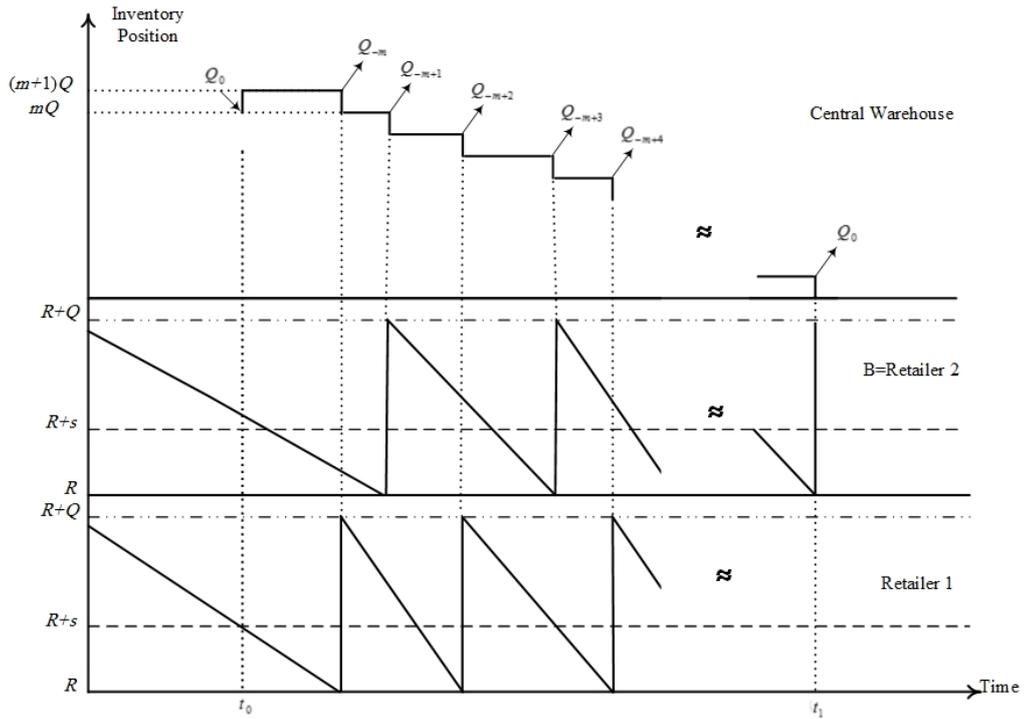

*Figure 2. An example of the system for two retailers*

Figure 2 represents the notations, note that for simplicity in our figures we use a continuous representation of demands to the retailers. Note that the $m + 1^{th}$ batch is $Q_0$; therefore, we use negative numbers for the $m$ batches received before this time (i.e. $-m$ to $-1$). The rest of the notations are as follows:

$Q_0$:  An arbitrary batch of size $Q$, chosen for investigation

$t_0$:  Ordering time of $Q_0$ by the central warehouse

$t_1$:  Ordering time of $Q_0$ by one of the retailers

$B$:  The retailer who orders $Q_0$

$t^+$:  A moment just after $t$

$t^-$:  A moment just before $t$

$k$:  Customer demand between $t_0$ and $t_1^+$

$j$:  An arbitrary unit of $Q_0$

$TC(m, R, s)$:   Average system cost per unit time

In what follows, we state the assumptions and inventory policy of this paper.

## 2.2. Assumptions

Similar to Moinzadeh (2002) and Afshar Sedigh et al. (2019) this study has the following assumptions:

1. Constant delivery time;
2. Poisson process with a constant, known, and identical rate of customer demand arrival;
3. Backlogged shortage;
4. The availability of online information about the retailer's inventory position and demand for the central warehouse;

We wish to state some reasons behind these assumptions. First, it is worth noting that Poisson arrival rate and constant delivery time extensively employed in literature; for instance, Axsäter (1990, 1993, 2000), Axsäter & Marklund (2008), and Forsberg (1995, 1997) are studies that considered both constant delivery time and Poisson/compound Poisson arrival rates. Also, several studies addressed retailers with identical demand rates to facilitate their modelling (e.g. Wang and Gunasekaran, 2017; Bradley, 2017; Nakade & Yokozawa, 2016; Fleischmann et al. 2020).

Backlogged shortage is a popular policy (e.g. see San-José, et al., 2018, 2019; Halim, 2017). Also, a known batch size along with a negligible ordering cost is widely used and acceptable in both literature and practice (e.g. see Axsäter, 1990, 1993, 2000; Axsäter & Marklund, 2008; Forsberg, 1995, 1997; Haji & Haji 2007; Svoronos & Zipkin, 1988, Deuermeyer & Schwarz, 1981; Moinzadeh, 2002). The reasons for such an approach include packaging or shipping requirements because of economies of scale in handling or shipping limitations, along with negligible shipping costs or benefiting from electronic commerce.

Finally, there are instances indicating that central warehouse had access to the information about retailers' demand activities and inventory status for more than two decades – Kurt Salmon Associates (1993,1997) reported employment of information technology in grocery industries, or Stalk et al. (1991) attributes Wal-Mart success to reasons such as detailed information sharing of customer's behavior. Therefore, it is acceptable to assume that the central warehouse has access to information about branches' inventory on hand, especially due to the availability of high-speed information exchange infrastructures and automated stores.

Having said the justifications, we restate Moinzadeh's (2002) policy; retailers use the $(R, Q)$ ordering policy, the central warehouse starts with $m$ initial batches ($m \geq 0$) and adopts the following ordering policy: Immediately after a retailer's inventory position reaches $R + s$, $(0 \leq s \leq Q - 1)$, a batch from an external supplier is ordered.

3. An Illustrative Example

Here we restate the example for two retailers from our earlier study (i.e. Section 3 of Afshar Sedigh et al., 2019) to obtain the central warehouse's inventory position. As can be seen in Figure 3, immediately after a retailer's inventory position reaches $R + s$, the central warehouse's inventory position is whether $(m + 1)Q$ or $(m + 2)Q$, considering inventory position of retailers at $t^-$ (Figure 7(A) and (B)).

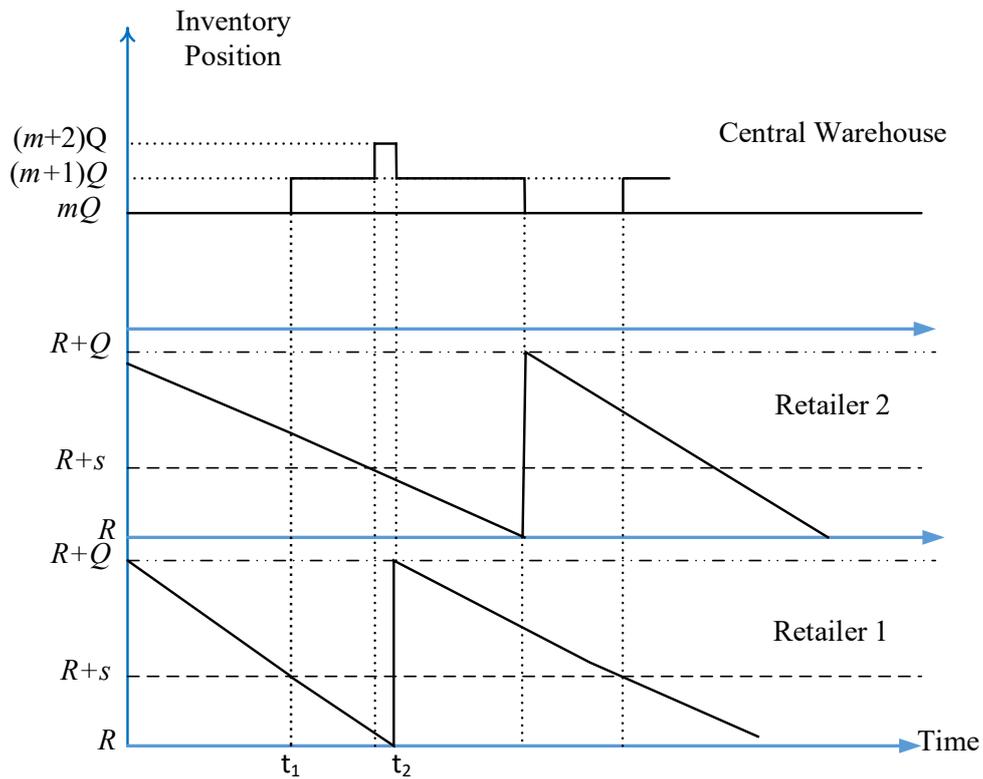

*Figure 3. The central warehouse inventory position*

For the rest of this section let $m = 3$, $Q = 4$, and $s = 2$; Figure 4(B) exemplifies the case that the retailer who initiates ordering $Q_0$ and the retailer who orders $Q_0$ are different. Overall, the retailers who initiates ordering $Q_0$ (i.e. the batch we are interested in) and the retailer to whom the batch is shipped are independent. For example, consider the situation that the inventory position of the retailer 1 is $R + s$ ($R + 2$) at $t_0$ (Figure 4(A) and (B)) – i.e. it initiates ordering $Q_0$. However, $Q_0$ can be shipped to Retailer 1 (Figure 4(A)) or Retailer 2 (Figure 4(C)). Henceforth, we address the cases that inventory position of Retailer 1 equals $R + 2$ at $t_0$.

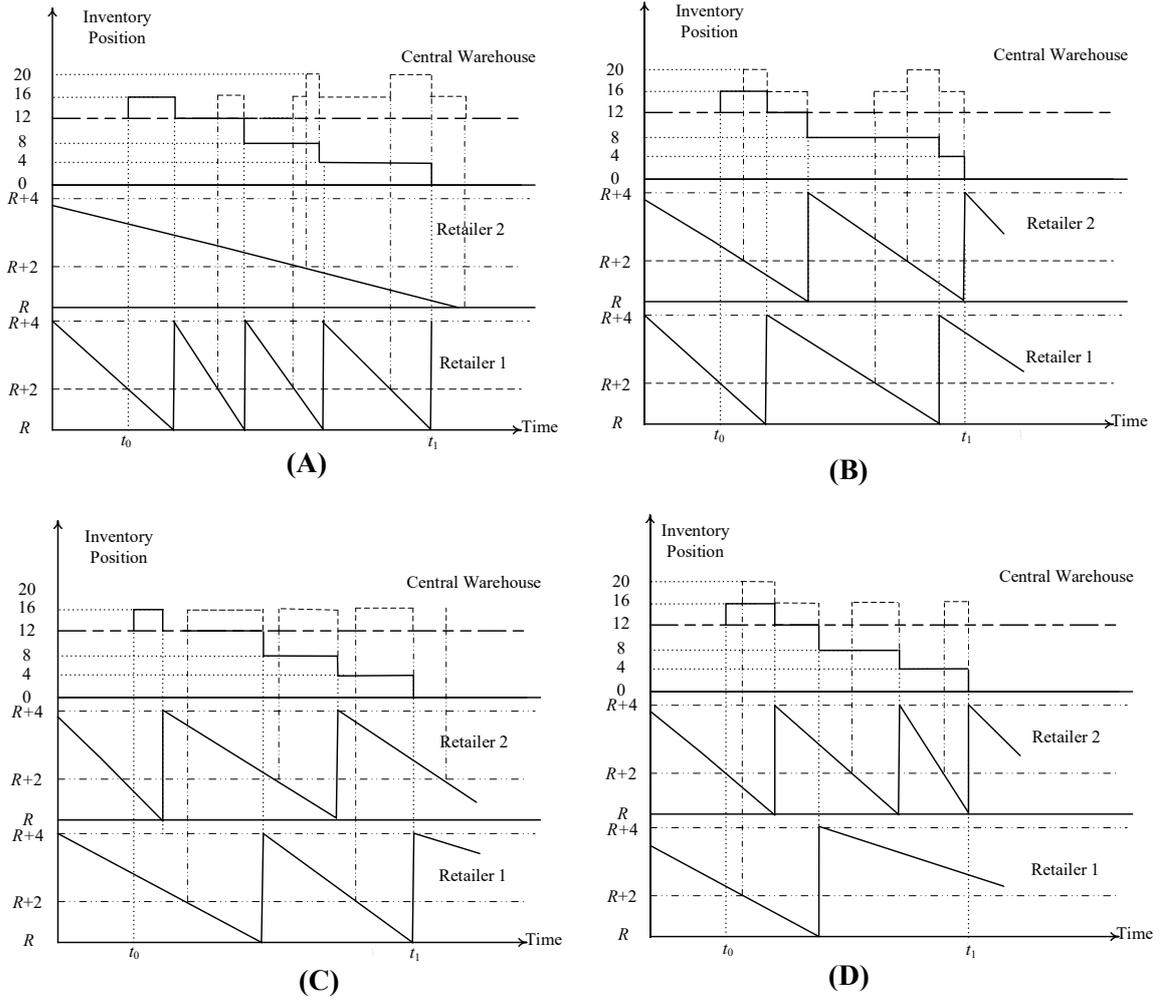

*Figure 4. Different possible combinations of retailers that cause the central warehouse to demand $Q_0$ and the retailer that demands $Q_0$ from the central warehouse when $I_w(t_0^-) = mQ$*

Let Retailer 1 order $Q_0$, an illustration of this system is presented in Figure 5. We know that between $t_0^+$ and $t_1^-$ three batches are ordered, and Retailer 1's inventory position equals $R + 1$ at $t_1^-$. Knowing that Retailer $i$ can order $b_i$ batches, so that $b_1 + b_2 = 3$, $b_i \geq 0, \forall i$, we want to calculate the probability of ordering these three batches by $l$ demands. Besides we know that, $l = l_1 + l_2$, and $l_r$ demands cause the retailer $r$ to order $b_r$ batches from the central warehouse. Let $l_r = b_r Q + u_r$, we need to obtain the probability that retailer $r$ receives $l_r$ demands of total $l$ demands during $t_1^- - t_0^+$ and the probability that these $l_r$ demands initiate $b_r$ batches ordering, i.e. $l_r = b_r Q + u_r$.

The probability of distributing $l$ demands between two retailers as $l = l_1 + l_2$ equals to $\binom{l}{l_1}\left(\frac{1}{2}\right)^l$ for identical demand rates. Moreover, the last demand should occur at Retailer 1. Thus, we eliminate the last demand. To do so, let $l' = l + 1$, $l'_1 = l_1 + 1$ and $l' = l'_1 + l_2$ and the probability is calculated as follows:

$$\frac{1}{2}\binom{l}{l_1}\left(\frac{1}{2}\right)^l = \frac{1}{2}\binom{l'-1}{l'_1-1}\left(\frac{1}{2}\right)^{l'-1} = \binom{l'-1}{l'_1-1}\left(\frac{1}{2}\right)^{l'} \tag{1}$$

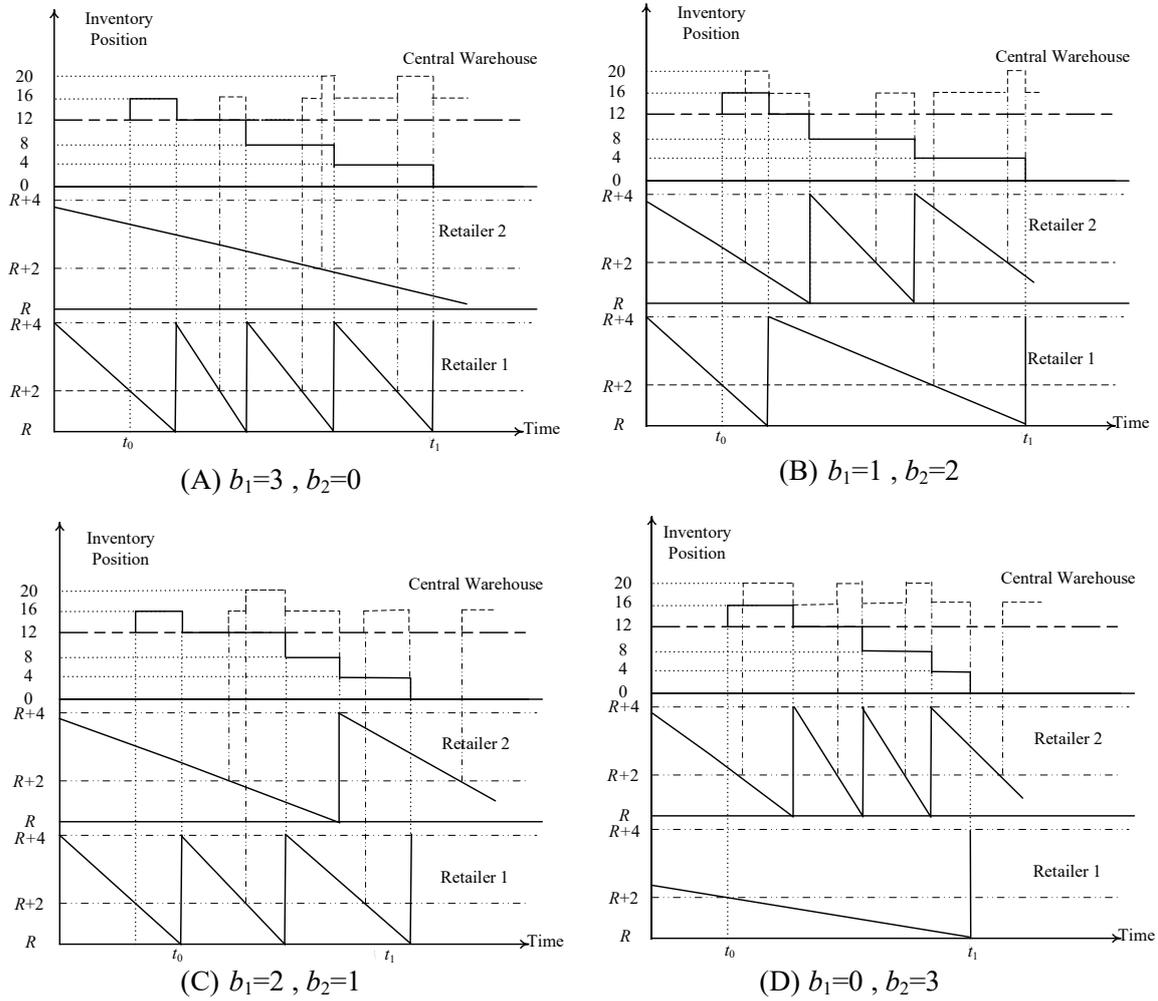

*Figure 5. Possible number of batches that retailers 1 and 2 can order when Retailer 1 causes the central warehouse to order $Q_0$ and consumes this batch itself*

Let $R + k'$ show the inventory position of Retailer 2 at $t_0$. In the example in question, we can calculate the upper and lower bounds of $l'$. To obtain the lower bound, inventory position of Retailer 2 at $t_1^-$ equals to $R + Q$; therefore,

$$l' \geq k' + 4(b_2 - 1) + 4b_1 + 2 \tag{2}$$

We know that $k' \geq 3$ and $b_1 + b_2 = 3$; therefore, $l' \geq 13$. Note that in Figure 5(A) the lower bound cannot be met, since $b_1 = 3$ and the lower bound is $4b_1 + 2 = 14$, i.e. $l' = 13$ is not feasible.

To obtain the upper bound, we note that Retailer 2's inventory position should be equal to $R+1$ at $t_1^-$, and we have:

$$l' \leq k' - 1 + 4b_2 + 4b_1 + 2 \tag{3}$$

On the other hand, we know that $k' \leq 4$ and $b_1 + b_2 = 3$; therefore, $l' \leq 17$.

Finally, $l' = l'_1 + l_2$ does not guarantee that all the conditions to be held. For example, let $l'_1 = 7$ and $l_2 = 9$; although $13 \leq l' = 16 \leq 17$, the inventory position of Retailer 1 at $t_1^-$ equals $R + 4$. On the other hand, let $l'_1 = 14$ and $l_2 = 3$, the inventory position of Retailer 1 at $t_1^-$ equals $R + 1$; but with probability $\frac{1}{2}$, $k' = 4$ which leads to a feasible solution. Therefore, we need to multiply (1) by some other probability functions to obtain the probability of demanding 3 batches by these $l'$ demands. In other words, for retailer $r$ we should calculate the probability of ordering $b_r$ batches by $l_r$ demand (note that $l_r = b_r Q + u_r$); we show this as $P(b_r Q + u_r \to b_r)$. Two illustrative computations for $P(b_2 Q + u_2 \to b_2)$ are presented in Figure 6. In Figure 6, we show the case that Retailer 2's inventory position is more than $R + 2$ at $t_0$ (i.e. it is either $R + 3$ or $R + 4$). We show the two different cases with distinctive lines. Figure 6(A) shows the calculation of $P(b_2 Q + u_2 \to b_2)$ for $u_2 = -1$ ($Q = 4$ and $b_2 = 3$), i.e. $P(11 \to 3)$. As can be seen, only one of the two cases is feasible the other case leads to ordering two batches; hence, the probability is $\frac{1}{2}$. Figure 6(B) shows the calculation of $P(b_2 Q + u_2 \to b_2)$ for $u_2 = 1$ and the same $Q$ and $b_2$, i.e. $P(13 \to 3)$. Here both cases are feasible; hence, the probability is 1.

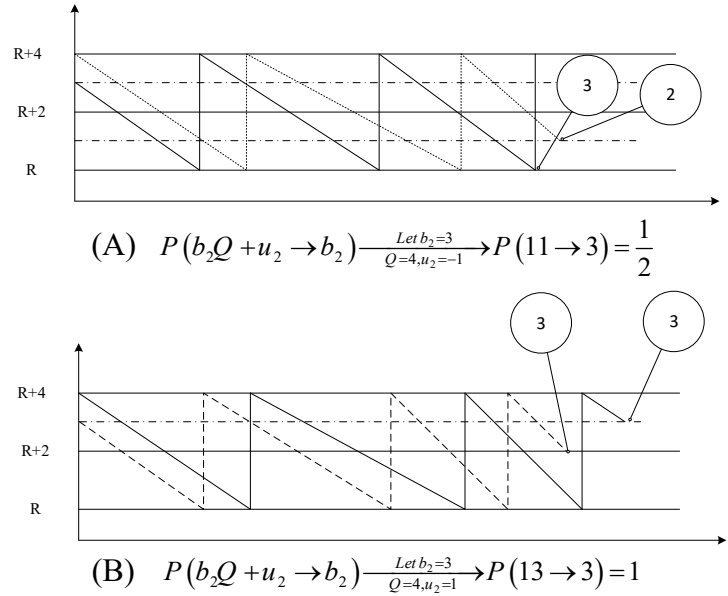

Figure 6. Illustrative computations for $P(b_2 Q + u_2 \to b_2)$ when $b_2 = 3$ and $u_2 = -1$ or $u_2 = 1$

## 4. Mathematical Model

This model extends Afshar Sedigh et al. (2019) for more than two retailers and obtains cost function of Moinzadeh (2002). In this section, we obtain the probability distribution and boundaries for customer demand. Finally, we obtain the mathematical model for this problem.

### 4.1. Probability distribution of customer demand ($k$)

The first step to obtain the probability distribution of the customer demand is to find the inventory positions of the central warehouse and the retailers at $t_0^-$. Note that at this moment (i.e. $t_0^-$) one of the retailers' inventory positions is $R + s$ (see $t_2$ in Figure 3). Furthermore, the inventory position at the central warehouse at this moment is a function of $Q, m,$ and $|A_0(t_0^-)|$, where $|A|$ is used to indicate the cardinality (number of members) of set $A$. In what follows, we obtain the inventory position at the central.

*4.1.1. The inventory position of central warehouse at $t_0^-$ when $s > 0$*

We wish to remind the readers that $A_0(t_0^-)$ is the set of retailers whose inventory positions at $t_0^-$ are less than or equal to $R + s$. Also, we know that inventory positions of retailers are uniformly distributed between $R + 1$ and $R + Q$ (Hadley and Whitin, 1963). We say the inventory system operates in state $i$ if $|A_0(t_0^-)| + 1 = i$. In order to obtain the probabilities associated with the system states, let $I_a(t_0^-)$ be the inventory position of retailer $a$ at $t_0^-$. Note that:

$$P(a \in A_0(t_0^-)) = P(R + 1 \leq I_a(t_0^-) \leq R + s) = \frac{s}{Q}$$

$$P(a \in A_1(t_0^-)) = P(R + s + 1 \leq I_a(t_0^-) \leq R + Q) = \frac{Q - s}{Q}$$
(4)

Thus, the probability that the system operates in state $i$ at $t_0$, i.e. $p(i, s)$, is:

$$P(i, s) = \binom{N-1}{i-1} \left(\frac{s}{Q}\right)^{i-1} \left(\frac{Q-s}{Q}\right)^{N-i} \quad ; \quad s > 0 \tag{5}$$

In state $i$, when $s > 0$, the inventory position of the central warehouse at $t_0^+$ is $(m + i)Q$. Furthermore, because the retailers are assumed to be identical, without losing generality the following sequence is assumed for the retailers (let $a$ be a retailer's number):

$$\begin{cases} a \in A_0(t_0^-) & \text{if } a \leq i - 1 \\ I_a(t_0^-) = R + s & \text{if } a = i \\ a \in A_1(t_0^-) & \text{if } a \geq i + 1 \end{cases} \tag{6}$$

In the abovementioned equation, for a system entering state $i$ at $t_0$, there are $i$ retailers with an inventory position with less than or equal to $R + s$. From the stated retailers one surely has an inventory position equal to $R + s$ (the one whom triggered the order by the central warehouse); we call it retailer $i$. For the rest of retailers, if their inventory position is less than or equal to $R + s$, they have a number from 1 to $i - 1$, else $i + 1$ to $N$.

Furthermore, we wish to remind the reader that we call the retailer who orders $Q_0$ and the time of ordering $Q_0$, $B$ and $t_1$, respectively. The inventory position of retailer $B$ can take one of the following values based on occurrence of three mutually exclusive events:

$$\begin{cases} R+1 \leq I_B(t_0) < R+s & if\ B \leq i-1 \\ I_B(t_0) = R+s & if\ B = i \\ R+s+1 \leq I_B(t_0) \leq R+Q & if\ B \geq i+1 \end{cases} \quad (7)$$

Let the system be in state $i$, where there are $m+i$ batches in the central warehouse at $t_0^+$. We know that the last customer demand for $Q_0$ had occurred at $t_1^-$ and the inventory position of $B$ decreased to $R$ at $t_1$.

Now we are ready to calculate the probability that $b$ batches are ordered by retailer $r$. The following two notations are used for this purpose:

$\mu_{r,l,B,s}^i$:     A random variable that represents the number of batches ordered by retailer $r$ when the system is in state $i$ and retailer $B$ orders $Q_0$ to fulfil $l$ demands

$\eta_{r,l,B,s}^i$:     A random variable that shows the number of batches ordered by the first $r$ retailers when the system is in state $i$ and retailer $B$ orders $Q_0$ after arriving $l$ demands to all retailers.

Note that the last customer demand is disregarded (it is assigned to retailer $B$). In this case, the schema of the inventory system is modified as shown in Figure 7. Therefore, the following recursive relation is used to determine the probabilities associated with different values of $\eta_{r,l,B,s}^i$.

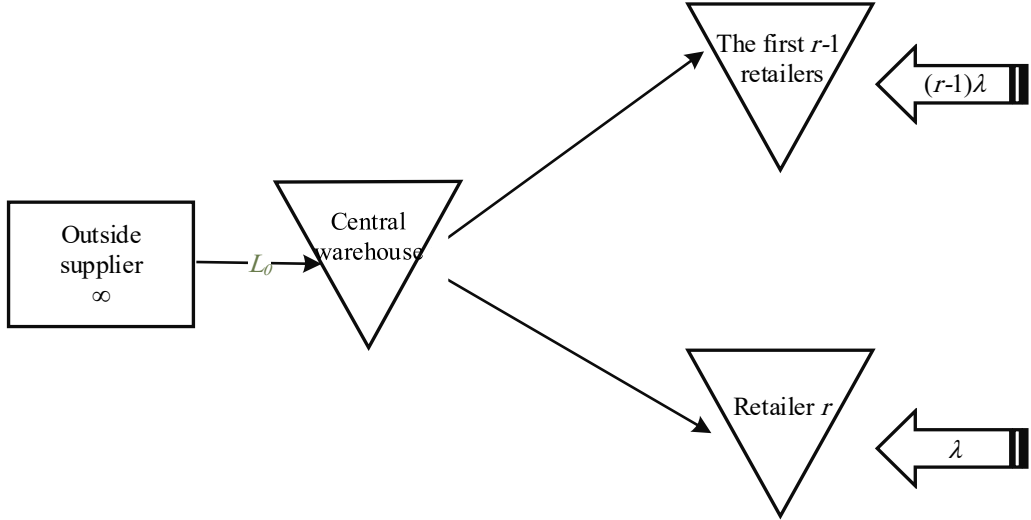

*Figure 7. System with reduced retailers*

$Pr(\eta^i_{r,l,B,s} = b)$

$$= \sum_{k=0}^{l} \sum_{b'=0}^{b} \binom{l}{k} \left(\frac{1}{r}\right)^k \left(\frac{r-1}{r}\right)^{l-k} \times Pr(\eta^i_{r-1,l-k,B,s} = b - b') \quad (8)$$

$$\times Pr(\mu^i_{r,k,B,s} = b')$$

where for $r = 1$ we define the following equation.

$$Pr(\eta^i_{1,l,B,s} = b) = Pr(\mu^i_{1,l-k,B,s} = b) \quad (9)$$

Because inventory position of each retailer is renewed every $Q$ customer demand (see Figure 3), evaluating system demands will be easier using $bQ + u$ instead of (i.e. we calculate $Pr(\mu^i_{r,bQ+u,B,s} = b))$. This probability depends on $s$, residual inventory ($u$), system state ($i$), and retailer ($r$). When the value of parameters stated earlier are given, the inventory position of retailer $B$ is defined in one of the intervals defined in (7).

*4.1.1.1 Calculating the probabilities for retailer who orders $Q_0$*

Here we study when $s > 0$ and $r = B$, retailer $B$ orders $b$ batches, when $bQ + u + I_B(t_0)$ equals $bQ + R + 1$ (the last demand should lead to ordering the last batch). Then using relation (7), it is possible to obtain acceptable values of $u$, based on which $Pr\left(\mu_{r,bQ+u,B,s}^{i} = b\right)$ is obtained by:

$$Pr\left(\mu_{r,bQ+u,B,s}^{i} = b\right) = \begin{cases} \frac{1}{s} & B < i, s > 0, r = B, 0 \leq u \leq s - 1, b \geq 0 \\ 1 & B = i, s > 0, r = B, u = s - 1, b \geq 0 \\ \frac{1}{Q-s} & B > i, s > 0, r = B, s \leq u < Q, b \geq 0 \end{cases} \quad (10)$$

*4.1.1.2 Calculating the probabilities for retailer who does not order $Q_0$*

When $s > 0$ and $r \neq B$, $u$ can take a value in one of the two intervals $0 \leq u \leq I_r(t_0) - R - 1$ and $I_r(t_0) - R \leq Q + u < Q$. Therefore, we have:

$$I_r(t_0) - Q - R \leq u \leq I_r(t_0) - 1 - R. \quad (11)$$

Then, using relations (7) and (11), the acceptable intervals of $u$ is obtained as follows:

$$\begin{cases} -Q + 1 \leq u \leq s - 1 & if\ r \leq i - 1 \\ 1 \leq u \leq s & if\ r = i \\ s + 1 - Q \leq I_B(t_0) \leq Q - 1 & if\ r \geq i + 1 \end{cases} \quad (12)$$

Relation (12) states that $-Q + 1 \leq u \leq s - 1$ for $s > 0$, $r \neq B$, and $r \leq i - 1$. It means that $u$ can take a v alue either in **(I)** $0 \leq u \leq s - 1$ or in **(II)** $-Q + 1 \leq u \leq -1$. The former occurs if $I_r(t_0) > R + u$ ($u$ should be less than $I_r(t_0) - R$). As $I_r(t_0)$ is uniformly distributed in $\{R + 1, \dots, R + s\}$, this event occurs with probability of:

$$Pr\left(\mu_{r,bQ+u,B,s}^{i} = b\right) = \frac{s-u}{s} \ ;\ B < i,\ s > 0,\ r \neq B,\ 0 \leq u \leq s,\ b \geq 0. \quad (13)$$

In order to obtain $Pr\left(\mu_{r,bQ+u,B,s}^{i} = b\right)$, for the interval defined in **(II)**, two cases are considered as stated in the following.

**(II-a)** $s - Q < u \leq -1$; here the customer demand is $u + Q$; hence, the interval can be modified as $s < u + Q \leq Q$. This demand surely causes a retailer to order, because it exceeds $I_r(t_0)$. Thus, the corresponding probability is 1. In other words, we have:

$$Pr(\mu^i_{r,bQ+u,B,s} = b) = 1 \quad ; \quad B < i, \; s > 0, \; r \neq B, \; s - Q < u < 0, \; b \geq 1. \tag{14}$$

**(II-b)** $-Q + 1 < u \leq s - Q$; in this situation $0 < u + Q \leq s$ and hence the customer demand causes a retailer to order if it is more than or equal to $I_r(t_0) - R$. Thus, the probability that $u + Q \geq I_r(t_0)$ is obtained as follows:

$$Pr(\mu^i_{r,bQ+u,B,s} = b) = \frac{Q+u}{s} \quad ; \quad B < i, \quad s > 0, \quad r \neq B, \quad -Q < u \leq s - Q, \quad b > 0 \tag{15}$$

Similarly, the following probabilities are obtained:

$$Pr(\mu^i_{r,bQ+u,B,s} = b) = \begin{cases} 1 + \dfrac{u}{Q-s}; & r > i, r \neq B, s > 0, s - Q < u \leq 0, b \geq 1 \\ 1; & r > i, r \neq B, s > 0, s - Q < u \leq 0, b \geq 0 \\ \dfrac{Q-u}{Q-s}; & r > i, r \neq B, s - Q < u \leq 0, b \geq 0 \\ 1 & r = i, r \neq B, s > 0, s - Q < u \leq 0, b \geq 1 \\ 1; & r = i, r \neq B, s > 0, s - Q < u \leq 0, b \geq 0 \end{cases} \tag{16}$$

*4.1.2. The inventory position of central warehouse at $t_0^-$ when $s = 0$*

When $s = 0$, the system surely operates in state 1 at $t_0$. In other words, without information sharing, the probability that the system operates in state $i$ at $t_0$ is obtained as:

$$p(i,s) = \begin{cases} \binom{N-1}{i-1} \left(\dfrac{s}{Q}\right)^{i-1} \left(\dfrac{Q-s}{Q}\right)^{N-i} & s > 0 \\ 1 & i = 1, s = 0 \\ 0 & otherwise \end{cases} \tag{17}$$

Note that $Pr(\mu^i_{r,bQ+u,B,s} = b)$ is meaningful when $B \geq 1$. Otherwise, the last demand must be considered in order to avoid negative customer demand, when $B = 1$ and $b = 0$. It is also obvious that at $t_0$, one batch is ordered by retailer 1 and the central warehouse, simultaneously. Thus, at $t_0^+$ the inventory position of the central warehouse is $m$ and that $I_1(t_0^+) = Q$.

The probability of demanding $b$ batches by retailer $r$ is as follow:

$$Pr(\mu_{r,bQ+u,B,s}^i = b) = \begin{cases} 1; & B = 1, s = 0, r = B, u = Q-1, b \geq 0 \\ \dfrac{1}{Q}; & B > 1, s = 0, r = B, 0 \leq u < Q, b \geq 0 \\ \dfrac{Q+u}{Q}; & r > 1, r \neq B, s = 0, -Q < u \leq 1, b \geq 1 \\ \dfrac{Q-u}{Q}; & r > 1, r \neq B, s = 0, 0 \leq u < Q, b \geq 0 \\ 1; & r = 1, r \neq B, s = 0, 0 \leq u < Q, b \geq 0 \end{cases} \quad (18)$$

Let the system be in state $i$, $m'(i,s)$ is defined as:

$$m'(i,s) = \begin{cases} m+i & s > 0 \\ m & s = 0 \end{cases} \quad (19)$$

therefore, we need to compute the probability of ordering $m'$ batches to fulfil exactly $k$ customer demands. As mentioned in computing $Pr(\mu_{r,bQ+u,B,s}^i = b)$, we didn't consider one customer demand to ensure that this customer demand was the last one. In our model, this demand will be assigned to retailer $B$, with probability of $\dfrac{1}{N}$. Therefore, the probability is calculated as:

$$\frac{1}{N} Pr(\eta_{N,l,B,s}^i = m'(i,s) - 1) \quad (20)$$

**4.2. Boundaries of Customer Demand**

Let us use $m'$ instead of $m'(i,s)$. Thus, the following lemma is proposed:

**Lemma 1**: When the system has $N$ retailers, from $m'$ batches at least $m' - N$ batches are ordered by demand sizes equal to $Q$.

**Proof:** this lemma states that at most $N$ batches can be ordered by less than $Q$ customer demand. Note that, when a retailer demands a batch, it meets its reorder point $(R)$. Furthermore, each retailer at most meets $R$ once with less than $Q$ customers order, and after that its inventory position equals $R + Q$. Consequently, the number of retailers $(N)$ gives the maximum number of the ordered batches with less than $Q$ customer demand.

Now let us introduce the following notations:

$lb(i, s, m)$: The lower bound of customer demand, when system starts by $m$ initial batches and is in state $i$ for a given $s$.

$ub(i, s, m)$: The upper bound of customer demand, when system starts by $m$ initial batches and is in state $i$ for a given $s$.

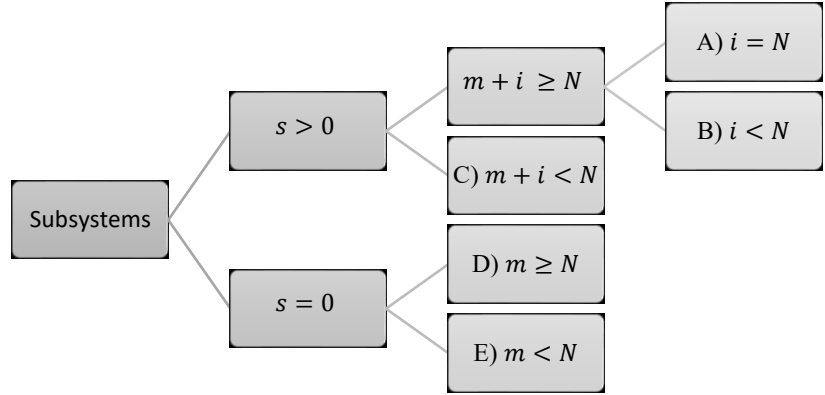

Figure 8. Five subsystems

We consider five subsystems as shown in Figure 5. This is based on differences in system behavior when $s > 0$ or $s = 0$ and the aforementioned lemma. The subsystems are stated in the following:

(A) $s > 0, m + I \geq N, i = N$;

(B) $s > 0, m + I \geq N, i < N$;

(C) $s > 0, m + I < N$;

(D) $s = 0, m < N$;

(E) $s = 0, m \geq N$.

3.2.1 Lower Bounds
When $s > 0$ and the system is in state $i$, for all subsystems we know that:

$$min\{I_r(t_0)\} = \begin{cases} R + 1 & r \leq i - 1 \\ R + s & r = i \\ R + s + 1 & r \geq i + 1 \end{cases} \quad (21)$$

Here $i - 1$ batches can be ordered by one customer demand to each one of the first $i - 1$ retailers. The other batches are divided into $m_1 = min(N - i + 1, m)$ and $m_2 = max(m - N + i, 0)$. First batch from the $m_1$ batches is ordered by $s$ customer demand to retailer $i$ and remained batches are ordered by $s + 1$ customer demand to the rest of $N - i$ retailers. It is obvious that the rest of $m_2$ batches are ordered by $m_2 Q$ customer demand to retailers. Therefore, the lower bounds when $s > 0$ is obtained as:

$$lb(i, s, m) = \begin{cases} i - 1 + s + m(s + 1) & m + i < N, s > 0 \\ i - 1 + s + (N - i)(s + 1) + (m + i - N)Q & m + i \geq N, s > 0 \end{cases} \quad (22)$$

When $s = 0$, at most $N - 1$ batches are ordered by one customer demand and remained batches, if there is any, are ordered by $Q$ customer demand. Therefore, the lower bounds are obtained as follows:

$$lb(i, s, m) = \begin{cases} m & m < N, s = 0 \\ N - 1 + (m - N + 1)Q & m \geq N, s = 0 \\ i - 1 + s + m(s + 1) & m + i < N, s > 0 \\ i - 1 + s + (N - i)(s + 1) + (m + i - N)Q & m + i \geq N, s > 0 \end{cases} \quad (23)$$

*3.2.2 Upper Bounds:*

When $s > 0$ and system is in state $i$, for all subsystems we know:

$$ub(i, s, m) = \begin{cases} max\{I_r(t_0)\} = R + s & r \leq i - 1 \\ I_r(t_0) = R + s & r = i \\ max\{I_r(t_0)\} = R + Q & r \geq i + 1 \end{cases} \quad (24)$$

To calculate the upper bound, it is assumed that $I_r(t_0)$s are at their maximum level for all retailers ($\forall r$). Also, we divide the system into $i < N$ and $i = N$. In the first case, there always is a retailer with $I_r(t_0) > R + s$. Therefore, all the $m + i$ batches can be ordered by one of the $N - i$ last retailers with $(m + i)Q$ demand. Moreover, we assume that the rest of retailers have $max\{I_r(t_0)\} - R - 1$ demand. For the second case, $i = N$, whereas $max_{\forall i}\{I_r(t_0)\} = s$, one of the batches is ordered by $s$ demand. The rest is the same as previous case. Finally, the only difference between $s = 0$ and $s > 0$ is in batches quantities. As stated earlier, when $s = 0$, there is $m$ batches in the central warehouse instead of $m + 1$.

$$ub(i, s, m) = \begin{cases} i(s - 1) + (N - i - 1)(Q - 1) + (m + i)Q & s > 0, i < N \\ (N - 1)(s - 1) + s + (m + i - 1)Q & s > 0, i = N \\ (N - 1)(Q - 1) + mQ & s = 0 \end{cases} \quad (25)$$

### 3.3. Final Model

Before representing the final model, note that we consider an arbitrary unit of $Q_0$, let it be $j^{th}$ unit of $Q_0$, with probability of $\frac{1}{Q}$. Furthermore, customer demand needed to order this unit between $t_0^+$ and $t_1$ equals $R + j$.

For computational simplicity, the following notation is introduced:

$f(i, B)$:   Number of retailers which are in the same situation as $r = B$, when system is in state $i$.

Because we have identical retailers, the problem is simplified by evaluating system costs, based on which retailer $i - 1$, $i$ or $i + 1$ orders $Q_0$ and then we multiply it by $f(i, B)$.

$$f(i, B) = \begin{cases} i - 1 & B = i - 1 \\ 1 & B = i \\ N - i & B = i + 1 \end{cases} \quad (26)$$

Given the aforementioned, the final model is obtained as follows:

$$TC(m, s, R) = \frac{\lambda}{N} \sum_{i=1}^{N} p(i, s) \sum_{B=i-1}^{i+1} \sum_{k=lb(i,s,m)}^{ub(i,s,m)} f(i, B) \times pr(\eta_{N,l,B,s}^i = m'(i, s) - 1)$$

$$\times \left( \gamma(k) + \frac{1}{Q} \sum_{j=1}^{Q} \Pi^{R+j}(k) \right) \quad (27)$$

### 4. Conclusion and discussion

In this paper, we obtained the exact model for a two-echelon inventory system with a central warehouse and a number of retailers with Poisson distributed demand which was an open question. In the model unfulfilled customer demand was backlogged. The batch size was given, for retailers and the central warehouse. Using conditional probabilities, we could derive the exact model for identical retailers. The model facilitates the optimization of Moinzadeh's (2002) policy and extending its model to more general scenarios (e.g. retailers with different delivery time).

Moinzadeh's (2002) policy has the benefit of simplicity for implementation that makes it a good candidate to utilize in the industries that does not want to invest in a complex policy such as the one proposed by Axsäter & Marklund (2008). This model can also be extended for retailers with unidentical demand rates; however, the extension would be more complex. Finally, it is worth noting that, although using a simple to implement policy may not reduce the costs the same as more flexible policies, such a policy enables more medium-sized businesses to utilize information sharing and reduce their costs in comparison with otherwise not sharing information in the supply chain.

**Appendix A**

Here a summary of Axsäter's (1990) cost functions is proposed. Let us define:

$\lambda_0$:    demand intensity at the central warehouse ($\sum_{i=1}^{N} \lambda_i$)

Note that in our paper $\lambda_0$ were equal to $N\lambda$.

The average holding cost in the central warehouse, $\gamma(S_0)$, is:

$$\gamma(S_0) = e^{-\lambda_0 L_0} \frac{h_0}{\lambda_0} \cdot \sum_{k=0}^{S_0-1} \frac{(S_0 - k)}{k!} (\lambda_0 L_0)^k, \quad S_0 > 0 \tag{A.1}$$

Or equally:

$$\gamma(S_0) = \frac{h_0 S_0}{\lambda_0} \cdot \left(1 - G_0^{S_0+1}(L_0)\right) - h_0 L_0 \left(1 - G_0^{S_0}(L_0)\right), \quad S_0 > 0 \tag{A.2}$$

For $S_0 \leq 0$ we have:

$$\gamma(S_0) = 0, \quad S_0 \leq 0 \tag{A.3}$$

The unit cost at retailer *I* for a given delay equal to *t* at the warehouse, is:

$$\pi_i^{S_i}(t) = e^{-\lambda_i(L_i+t)} \frac{h_i+\beta_i}{\lambda_i} \cdot \sum_{k=0}^{S_i-1} \frac{(S_i-k)}{k!} \lambda_i^k (L_i+t)^k + \beta_i \left(L_i + t - \frac{S_i}{\lambda_I}\right) \quad (A.4)$$

For $S_i \leq 0$, average holding and shortage cost in retailer *I*, $\Pi_i^{S_i}(S_0)$ is:

$$\Pi_i^{S_i}(S_0) = G_0^{S_0}(L_0)\beta_i L_0 - G_0^{S_0+1}(L_0)\beta_i \frac{S_0}{L_0} + \beta_i(L_i - \frac{S_i}{L_i}) \quad , S_i \leq 0, S_0 > 0 \quad (A.5)$$

And:

$$\Pi_i^{S_i}(S_0) = \beta_i(L_i + L_0 - \frac{S_i}{L_i} - \frac{S_0}{L_0}) \quad , S_i \leq 0, S_0 \leq 0 \quad (A.6)$$

And for $S_i > 0$ it is possible to obtain the following recursive equation:

$$\Pi_i^{S_i}(S_0 - 1) = \frac{\lambda_i}{\lambda_0} \Pi_i^{S_i-1}(S_0) + \frac{\lambda_0-\lambda_i}{\lambda_0} \Pi_i^{S_i}(S_0) + \frac{\lambda_i}{\lambda_0}\left(1 - G_0^{S_0}(L_0)\right)\left(\pi_i^{S_i}(0) - \pi_i^{S_i-1}(0)\right) \quad , S_0 > 0 \quad (A.7)$$

And for $S_0 \leq 0$ we have:

$$\Pi_i^{S_i}(S_0 - 1) = \frac{\lambda_i}{\lambda_0} \Pi_i^{S_i-1}(S_0) + \frac{\lambda_0-\lambda_i}{\lambda_0} \Pi_i^{S_i}(S_0) \quad , S_0 \leq 0 \quad (A.8)$$

And the $\Pi_i^0(S_0)$ can be obtained by (A.5) and (A.6).

Finally, for sufficiently large values of $S_0$ the delay at the warehouse is equal to 0; therefore, we have:

$$\Pi_i^{S_i}(S_0) \approx \pi_i^{S_i}(0) \quad (A.9)$$

This approximation is asymptotically exact. To determine an initial large value of $S_0$, let the probability that the warehouse can deliver without a delay is smaller than ε:

$$G_0^{\overline{S_0}}(L_0) < \varepsilon \quad (A.10)$$

Where ε is a small positive number.

**References**

Afshar Sedigh, A. H., Haji, R., & Sajadifar, S. M. (2019). Cost function and optimal boundaries for a two-level inventory system with information sharing and two identical retailers. *Scientia Iranica. Transaction E, Industrial Engineering*, *26*(1), 472-485.

Axsäter, S. (1990). Simple solution procedures for a class of two-echelon inventory problems. *Operations research*, *38*(1), 64-69.

Axsäter, S. (1993). Exact and approximate evaluation of batch-ordering policies for two-level inventory systems. *Operations research*, *41*(4), 777-785.

Axsäter, S. (2000). Exact analysis of continuous review (R, Q) policies in two-echelon inventory systems with compound Poisson demand. *Operations research*, *48*(5), 686-696.

Axsäter, S., & Marklund, J. (2008). Optimal position-based warehouse ordering in divergent two-echelon inventory systems. *Operations Research*, *56*(4), 976-991.

Axsäter, S., & Rosling, K. (1993). Installation vs. echelon stock policies for multilevel inventory control. *Management Science*, *39*(10), 1274-1280.

Bradley, J. R. (2017). The Effect of Distribution Processes on Replenishment Lead Time and Inventory. *Production and Operations Management*, *26*(12), 2287-2304.

Chen, F., & Zheng, Y. S. (1994). Evaluating echelon stock (R, nQ) policies in serial production/inventory systems with stochastic demand. *Management Science*, *40*(10), 1262-1275.

Chen, F., & Zheng, Y. S. (1998). Near-optimal echelon-stock (R, nQ) policies in multistage serial systems. *Operations research*, *46*(4), 592-602.

Chen, M. C., Yang, T., & Yen, C. T. (2007). Investigating the value of information sharing in multi-echelon supply chains. *Quality & quantity*, *41*(3), 497-511.

Cho, D. W., & Lee, Y. H. (2013). The value of information sharing in a supply chain with a seasonal demand process. *Computers & Industrial Engineering*, *65*(1), 97-108.

Deuermeyer, B. L., & Schwarz, L. B. (1979). *A model for the analysis of system service level in warehouse-retailer distribution systems: the identical retailer case*. Institute for Research in the Behavioral, Economic, and Management Sciences, Krannert Graduate School of Management, Purdue University.

Fleischmann, M., Kloos, K., Nouri, M., & Pibernik, R. (2020). Single-period stochastic demand fulfillment in customer hierarchies. *European Journal of Operational Research*.

Forrester, J. W. (1958). Industrial Dynamics. A major breakthrough for decision makers. *Harvard business review*, *36*(4), 37-66.

Forsberg, R. (1995). Optimization of order-up-to-S policies for two-level inventory systems with compound Poisson demand. *European Journal of Operational Research*, *81*(1), 143-153.

Forsberg, R. (1997). Exact evaluation of (R, Q)-policies for two-level inventory systems with Poisson demand. *European journal of operational research*, *96*(1), 130-138.

Hadley, G., & Whitin, T. M. (1963). *Analysis of inventory systems*. Englewood Cliffs, N.J: Prentice-Hall.

Halim, M. A. (2017). Weibull distributed deteriorating inventory model with ramp type demand and fully backlogged shortage. *Asian Journal of Mathematics and Computer Research*, 148-157.

Haji, R., & Haji, A. (2007). One-for-one period policy and its optimal solution. *Journal of Industrial and Systems Engineering*, *1*(2), 200-217.

Haji, R., & Sajadifar, S. M. (2008). Deriving the exact cost function for a two-level inventory system with information sharing. *Journal of Industrial and Systems Engineering*, *2*(1), 41-50.

Kurt Salmon Associates. (1993). *Efficient Consumer Response:[ECR]; enhancing consumer value in the grocery industry*. Research Department Food Marketing Institute, Kurt Salmon Associates, Inc., Washington, Dc (1993).

Kurt Salmon Associates. (1997). *Quick Response: Meeting Customer Need*. Kurt Salmon Associates, Atlanta, GA.

Li, M., & Simchi-Levi, D. (2020). The Web Based Beer Game-Demonstrating the Value of Integrated Supply-Chain Management. In *MIT forum for Supply Chain Management Web site. Retrieved July*.

Li, X., & Wang, Q. (2007). Coordination mechanisms of supply chain systems. *European journal of operational research*, *179*(1), 1-16.

Liu, C., Huo, B., Liu, S., & Zhao, X. (2015). Effect of information sharing and process coordination on logistics outsourcing. *Industrial Management & Data Systems*.

Liu, C., Xiang, X., & Zheng, L. (2020). Value of information sharing in a multiple producers–distributor supply chain. *Annals of Operations Research*, *285*(1-2), 121-148.

Moinzadeh, K. (2002). A multi-echelon inventory system with information exchange. *Management science*, *48*(3), 414-426.

Nakade, K., & Yokozawa, S. (2016). Optimization of two-stage production/inventory systems under order base stock policy with advance demand information. *Journal of Industrial Engineering International*, *12*(4), 437-458.

Ojha, D., Sahin, F., Shockley, J., & Sridharan, S. V. (2019). Is there a performance tradeoff in managing order fulfillment and the bullwhip effect in supply chains? The role of information sharing and information type. *International Journal of Production Economics*, *208*, 529-543.

Sajadifar, S. M., & Haji, R. (2007). Optimal solution for a two-level inventory system with information exchange leading to a more computationally efficient search. *Applied mathematics and computation*, *189*(2), 1341-1349.

San-José, L. A., Sicilia, J., & Alcaide-López-de-Pablo, D. (2018). An inventory system with demand dependent on both time and price assuming backlogged shortages. *European Journal of Operational Research*, *270*(3), 889-897.

San-José, L. A., Sicilia, J., González-De-la-Rosa, M., & Febles-Acosta, J. (2019). Analysis of an inventory system with discrete scheduling period, time-dependent demand and backlogged shortages. *Computers & Operations Research*, *109*, 200-208.

Simchi-Levi, D., Kaminsky, P., Simchi-Levi, E., & Shankar, R. (2008). *Designing and managing the supply chain: concepts, strategies and case studies*. Tata McGraw-Hill Education.

Simchi-Levi, D., & Zhao, Y. (2007). *Three generic methods for evaluating stochastic multi-echelon inventory systems*. Working paper, Massachusetts Institute of Technology, Cambridge.

Simchi-Levi, D., & Zhao, Y. (2012). Performance evaluation of stochastic multi-echelon inventory systems: A survey. *Advances in Operations Research*, *2012*.

Stalk, G., Evans, P., & Shulman, L. E. (1992). Competing on capabilities: The new rules of corporate strategy. *Harvard business review*, *70*(2), 57-69.

Svoronos, A., & Zipkin, P. (1988). Estimating the performance of multi-level inventory systems. *Operations Research*, *36*(1), 57-72.

Wang, G., & Gunasekaran, A. (2017). Operations scheduling in reverse supply chains: Identical demand and delivery deadlines. *International Journal of Production Economics*, *183*, 375-381.

Zhou, H., & Benton Jr, W. C. (2007). Supply chain practice and information sharing. *Journal of Operations management*, *25*(6), 1348-1365.